\documentclass[article]{aastex631}

\shorttitle{FUOR flares in the clumpy accretion model}
\shortauthors{Demidova \& Grinin}

\begin{document}

\title{3-D SPH simulations of the FUOR flares in the clumpy accretion model}
\correspondingauthor{Tatiana V. Demidova}
\email{proxima1@list.ru}

\author[0000-0001-7035-7062]{Tatiana V. Demidova}
\affiliation{Crimean Astrophysical Observatory,p. Nauchny, Bakhchisaray, Crimea, 298409, Russia}

\author[0000-0001-8923-9541]{Vladimir P. Grinin}
\affiliation{Pulkovo Observatory of the Russian Academy of Sciences, Pulkovskoje Avenue 65, St. Petersburg 196140, Russia}
\affiliation{V.V. Sobolev Astronomical Institute, St. Petersburg University, Petrodvorets, Universitetskiy p. 28, St. Petersburg, 198504, Russia}

\begin{abstract} One of the early hypotheses about the origin of FUOR outbursts explains them by the fall of gas clumps from the remnants of protostellar clouds onto protoplanetary disks surrounding young stars~{ \citep{1985ApJ...299..462H}}. To calculate the consequences of such an event we make 3D hydrodynamic simulations by SPH method. It is shown that the fall of the clump on the disk in the vicinity of the star actually causes a burst of the star's accretion activity, resembling in its characteristics the flares of known FUORs. In the region of incidence, an inhomogeneous gas ring is formed, which is inclined relative to the outer disk. During several revolutions around the star, this ring combines with the inner disk and forms a tilted disk. In the process of evolution, the inner disk expands, and its inclination relative to the outer disk decreases. After $100$ revolutions, the angle of inclination is a few degrees. This result is of interest in connection with the discovery in recent years of protoplanetary disks, the inner region of which is inclined relative to the outer one. Such structures are usually associated with the existence in the vicinity of a star of a massive body (planet or brown dwarf), whose orbit is inclined relative to the plane of the disk. The results of our modeling indicate the possibility of an alternative explanation for this phenomenon.
\end{abstract}


\section{Introduction}
FU Ori stars are a rather small family of young eruptive stars. To date, only about 30 such objects are known~\cite[including candidates for FUORs,][]{2014prpl.conf..387A}. Flares of classical FUORs, such as V1057 Cyg and FU Ori, are characterized by a rapid increase in brightness in the initial phase of the outburst followed by a slow decline lasting decades~\cite[see, e.g.][]{1977ApJ...217..693H, 2013MNRAS.434...38K, 2021ApJ...917...80S}.  The accretion rate increases by several orders of magnitude and reaches $10^{-5}$--$10^{-3} M_\odot yr^{-1}$. At the same time, the rise time is less than $10$ years, and the outburst itself lasts from several to hundreds or more years.
It is assumed that such events can occur repeatedly during the evolution of a young star.  

There are several hypotheses about the origin of FUORs flares. Their detailed analysis is given in reviews~\citet{1996ARA&A..34..207H, 2014prpl.conf..387A,2022arXiv220311257F}. 
It is believed that the flares is caused by a sharp increase in the rate of accretion onto a young object. This leads to the fact that near the star the accretion disk is heated due to viscous friction to a temperature of the order of $10^4$~K, and this region becomes the dominant source of the object's optical radiation. This is what causes the flare. Direct evidence in support of such a scenario has recently been obtained by~\cite{2021A&A...646A.102L} using interferometric observations of FU Ori in the near infrared region of the spectrum.

The question is, what is the reason for the sharp increase in the accretion rate? This issue has been the subject of numerous studies, the conclusions of which can be briefly summarized as follows: an increase in the accretion rate may be the result of the development of instabilities in the circumstellar disk~\cite[see reviews][and literature cited there]{1996ARA&A..34..207H, 2014prpl.conf..387A}. One of the first hypotheses about the origin of FUOR flare was the idea that a sudden increase in the accretion rate of gas onto a young star could be the result of the falling onto the circumstellar disk of a massive fragment from the remnants of a protostellar cloud~\cite{1996ARA&A..34..207H}.   

 In~\citet{2011ApJ...736...53O, 2014ApJ...797...32P, 2018MNRAS.474.1176J} it was shown that the outbursts of the luminosity of stars in clusters can be associated with a late infall of matter onto it. The capture of cloudlet by a star with a disk was modeled in~\citet{2018MNRAS.475.2642K,2019A&A...628A..20D,2021A&A...656A.161K}. It was shown that the outburst luminosity is an order of magnitude lower than that observed for FU Ori. Also, the accretion rate during the passage of the cloudlet around the star did not exceed $10^{-9} M_\odot yr^{-1}$. But such a process can trigger a gravitational instability in the disk, which will contribute to the rapid transfer of angular momentum to the star. Which, in turn, can trigger an outburst of the accretion rate~\citep{2018MNRAS.475.2642K}. However, the outburst luminosity is an order of magnitude lower than that observed for FU Ori. Similar calculations, taking into account the magnetic field of the disk and cloud, were carried out in~\citet{2022ApJ...941..154U}. It was shown that if the size of the cloudlet is less or equal to the thickness of the disk, then the magnetic field slows down the rotation of the falling matter, but in the case of a larger cloudlet, the matter can move at super-Keplerian velocities.

 Another mechanism capable of causing a strong flare in the accretion rate is the flyby of another star near a star with a protoplanetary disk~\citep{2008A&A...492..735P,2010MNRAS.402.1349F}.
It was shown that this mechanism can create a flare of the accretion rate up to $10^{-3} M_\odot yr^{-1}$, but the increase in brightness takes tens of years. The case when the intruder star also has a protoplanetary disk was considered in the papers~\citep{ 2019MNRAS.483.4114C,2020MNRAS.491..504C}. The accretion rate in these calculations did not exceed $10^{-6} M_\odot yr^{-1}$, however, the time of an increase in the accretion rate by an order of magnitude occurred in $10$ years or less. 

In~\citet{2022MNRAS.510L..37B} similar calculations were performed taking into account the transfer of radiation. The authors showed that in the case of an incoming star, the accretion rate can imitate the FUOR flare in terms of the amplitude and time scales of the flare development. A study of a close flyby of two stars with disks was carried out in~\citet{2022MNRAS.517.4436B}. It has been shown that the presence of a disk in the incoming star increases the amplitude of the accretion rate in the target star. In this case, the retrograde approach leads to an increase in the accretion rate of the target star by more than $10^{-5}M_\odot yr^{-1}$.

An analysis of the three mechanisms leading to an increase in the accretion rate was carried out in~\citet{2021A&A...647A..44V}. The authors showed that the burst of the accretion rate during magneto-rotational instability is a less energetic process than the fall of the clump and the passage of a star. In addition, the last two mechanisms give an asymmetric accretion rate change profile. In the case of the impact of the clump on the disk, the time of increase in the accretion rate is about $10$ years.

A two-dimensional model of migration of massive gas clumps formed at the periphery of disks as a result of gravitational instability was considered in~\citet{2005ApJ...633L.137V} . It was shown that in the process of motion in the disk, the clumps lose part of their angular momentum and matter, interacting with density waves, and dissipate at some distance from the star, causing an increase in the accretion rate. {In such a model, a series of outbursts of accretion occurs (for several clumps falling), in addition, the orbital motion of the clump leads to a variation in the accretion rate~\citep{2021A&A...647A..44V}.}

 In~\citet{2012ApJ...750...30B, 2017A&A...608A.107V} it has been shown that some clumps can be ejected from a massive disk into interstellar space. Such wandering objects can be a source of clumpy accretion in young star clusters. The collision of such a clump with a protoplanetary disk can lead to a number of observational phenomena, including an outburst of the accretion rate. Therefore, it is of interest to study the consequences of such a process. 

In the paper, we study by the SPH method a three-dimensional model of dynamic relaxation of a circumstellar disk after a gas and dust clump falls on it. In our previous paper ~\citet{2022ApJ...930..111D} (Paper I) it was shown that when a massive clump (with a mass of about $0.1$ $M_{J}$) falls on the disk, spiral density waves arise in the disk, which pass into ring structures. Such structures are observed in disk images obtained with the ALMA interferometer. Below we will continue modeling clumpy accretion and consider the possibilities of this model in terms of creating outbursts of stellar accretion activity, similar to flares of FUORs.
\newpage
\section{Simulation}
\subsection{Initial conditions}
A model consisting of a star, with the parameters of the Sun $M_{\ast} = M_{\odot}$, $R_{\ast}=R_{\odot}$, $T_{\ast} = 5780$ K is considered. It is surrounded by a gaseous disk of mass $M_{disk}
= 0.01M_{\odot}$. At the initial moment of time, the disk matter is distributed azimuthally symmetrically within the inner ($r_{in} = 0.5$~au) and outer ($r_{out} = 50$~au) radii. The Cartesian coordinates of the particles $(x,y,z)$ are randomly assigned so that the distribution probability density corresponds the initial density of matter of the disk, which is described by the law
\begin{equation}
\rho(r,z,0)=\frac{\Sigma_0}{\sqrt{2\pi}H(r)}\frac{r_{in}}{r}e^{-\frac{z^2}{2H^2(r)}},
\label{Eq:rho}
\end{equation}
where $r=\sqrt{x^2+y^2}$ is cylindrical radius, $\Sigma_0$ is the parameter determined by the total mass of the simulated disk. The semi-thickness of the disk is determined by the following formula
\begin{equation}
H(r)=\sqrt{\frac{\kappa T_{mid}(r) r^3}{GM_{\ast} \mu m_H}},
\end{equation} 
where $\kappa$, $G$, and $m_H$ are the Boltzmann constant, the gravitational constant, and the mass of the hydrogen atom. The average molecular weight is $\mu=2.35$~\citep{1994A&A...286..149D}.

 Particle velocities are given as Keplerian at a certain distance from the star ($R=\sqrt{r^2+z^2}$)\footnote{Here and below $r$ is cylindrical radius and $R$ is spherical radius}, where $V_k(R)=\sqrt{\frac{GM_\ast}{R}}$ is the Keplerian velocity. Thus, the particle at the initial moment of time has the velocity coordinates $(-V_k(R) sin(\phi),V_k(R) cos(\phi),0)$, where $tan(\phi)=\frac{y}{x}$. The disc material moves counterclockwise.

The disk is assumed to be isothermal in the vertical direction, which is justified by its small thickness compared to the radius. The temperature distribution law, which determines the speed of sound in the disk, is described as follows
\begin{equation}
T_{mid}(r)=\sqrt[4]{\frac{\Gamma}{4}}\sqrt{\frac{R_{\ast}}{r}}T_{\ast},
\label{temp}
\end{equation}
where $\Gamma=0.05$ \citep{1997ApJ...490..368C,2004A&A...421.1075D}.

 Gas-dynamic calculations were performed without taking into account the dust component of the clump and protoplanetary disk.

For calculating images of a protoplanetary disk, the dust part was added to the gas component of the disk. The ratio of the mass of dust to the mass of gas in the disk was $0.01$ as the average in the interstellar medium. It was assumed that dust particles $1$, $10$, $100 \mu $m and $1$ mm in size are well mixed with gas and distributed according to the law $\frac{dn(s)}{ds}\propto s^{- 3.5}$, where $n$ is the concentration, s is the size of the dust grain \citep{1969JGR....74.2531D}. The dust was magnesium-iron silicates~\citep{1995A&A...300..503D}.

\subsection{Discrete clump-disk collision approximation}
 In Paper I, the impulse approximation of the clump disk collision was considered: the gaseous clump was added to the disk at the time t = 0 as a density perturbation in the form of a truncated sector of the disk with an azimuth angle of 30$^\circ$. In this paper, the clump was divided into $N$ parts, which were sequentially added to the disk as a density perturbation. Each clump element was added to the disk region bounded by the inner radius $r_0$ with a width $dr$, and the azimuth angle was $\Delta \phi= 30^\circ/N$ with the symmetry axis along the negative part of the $x$ axis. The clump moves in the same direction as the matter of the disk with the velocity $\displaystyle V(R) = L\cdot V_k(R)$, where  $L=0.8$ ($L = 0.66$ for the extreme model) is a dimensionless quantity. The velocity vector has a residual inclination to the disk plane $\displaystyle sin(I) = \frac{V_z(R)}{V(R)}$. Clump particle with coordinates $(x,y,z)$ at the initial moment of time has velocity coordinates $(-V(R) sin(\phi) cos(I),V(R) cos(\phi) cos(I), V(R) sin(I))$.
 
 The residual inclination angle $I$ depends on the initial angle of falling and the amount of kinetic energy that is spent on heating the disk in the region of incidence. Below we consider a model in which $I = 30^\circ$ for the usual models and $I = 45^\circ$ for the extreme model. The next element of the clump was added after a time interval $\Delta t$, during which the azimuthal angle of a point in the disk plane $r_0+\frac{dr}{2}$ au will change to $\Delta \phi$ during Keplerian rotation. In calculations, $N = 6$ was taken. In this way, we simulated a fall on the disk of a clump in the form of a gas stream of finite size. It was assumed that the composition of the clump is identical to substance of the protoplanetary disk. The initial density of the clump was determined by its mass and differed from the density of the protoplanetary disk.
 
The clump mass ($m_C$) varied from $0.1$ to $3$ Jupiter masses ($M_J$). Whereas the mass of the disk contained within the radius $r = 7$~au (the case $(r_0+dr)<7$~au was considered) is approximately $0.3 M_J$. Thus, the perturbation mass is comparable to or exceeds the mass of the central part of the disk (the initial density exceeds the density in the disk by $10$--$100$ times). In this case, the density of the clump matter at the initial moment of time significantly exceeds the density of the disk matter. Thus, the properties of the matter in the region of the fall of the clump will be determined by the characteristics of the clump. 

When the clump falls on the disk, part of its kinetic energy is spent on heating the gas at the point of impact. In the model under consideration, the thickness of each portion of the perturbation along the azimuthal angle and along the height $\sim H$.  According to~\citet{2017A&A...605A..30M}, the characteristic thermal relaxation time for low-frequency modes $\lambda/H=2$ for disk mass $0.01M_\odot$ is comparable to the time of the orbital period at a distance of about $3$~au. This value can be smaller with a decrease portion of dust in a gaseous medium. Also, relaxation in the vertical direction proceeds much faster than in the radial direction. Therefore, the thermal relaxation time can be less than the orbital period. This makes it possible to consider the process of dynamic relaxation of the disk using the temperature regime calculated in the stationary approximation.      

\subsection{Method of calculation}
\label{sec:method}
As in Paper I, the integration of the equation of motion was performed by the SPH (smoothed particles hydrodynamics) method using the Gadget-2 code~\citep{2001NewA....6...79S, 2005MNRAS.364.1105S} modified by us~\citep{2016Ap.....59..449D}.
It was assumed that, in addition to gas-dynamic forces, the equation of gas motion includes forces due to artificial viscosity (proportional to the speed of sound), gravitation of the star, and self-gravity of the disk. The calculations were performed in a non-rotating Cartesian coordinate system with the star at the origin. 

The temperature distribution law, defined by the Eq.~\ref{temp}, was preserved throughout the calculations, which were performed in the approximation of local thermodynamic equilibrium. Thus, the equation of state of the disk matter was defined as $P(r,z,t)=c^2(r)\rho(r,z,t)$, where $P$ is the local pressure at time $t$, and $c$ is the speed of sound at a given point on the disk. The disk model described above is relaxed for $600$ years. Then a gas clump with the parameters indicated above above is added into the disk. After that, the system evolved for another $2000$ years.

The gaseous protoplanetary disk was modeled using $2.5\times10^6$ SPH particles, and the clump of matter at the initial moment of time contained $\sim5\times10^5$ particles. Particles that went beyond $150$~au were considered to have left the system. Particles that came closer to center of disk than $0.5$~au were considered to have accreted on the star.

 The accretion rate was calculated in increments of $1-5$ years, then averaged over three points and smoothed using a three-point moving average. To analyze the behavior of the accretion rate, the accretion rate gradient was used. The beginning of the outburst denotes the moment when the accretion rate gradient begins to increase. The duration of the flare is determined by the difference between the times of the minimum and maximum of the accretion rate gradient, and the maximum of the flare is determined by the time moments when the gradient is zero.

The simulated area was divided into $200\times30\times90$ cells in spherical coordinates ($R,\theta,\phi$), in which the average values of SPH particle density were determined. The RADMC-3D~\citep{2012ascl.soft02015D} code was used for 3D radiative transfer calculations. We use $10^9$ photons in direct radiation and scattering calculations. Dust opacity for Magnesium-iron silicates~\citep{1995A&A...300..503D} is calculated by using the Mie code~\citep{1998asls.book.....B} included in  RADMC-3D package.

\section{Results}
\subsection{The behavior of matter in the disk}
\begin{figure}
\centering
\includegraphics[width=0.7\textwidth]{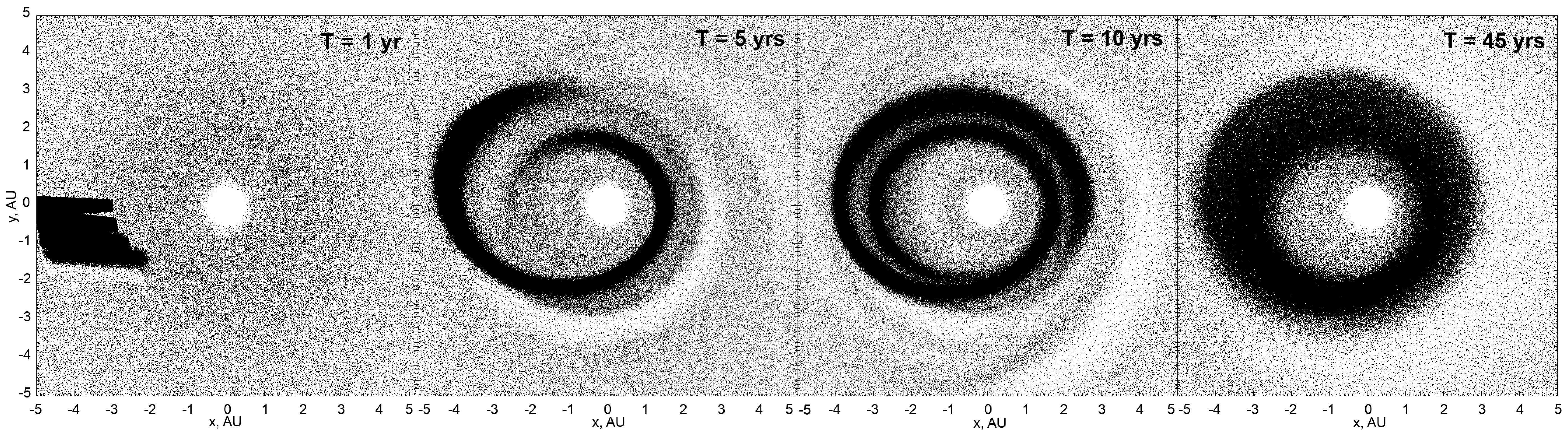}
\caption{\label{fig:disk} Projection of particles onto the disk plane after $1$~yr (when all parts of the clump are added to the disk) and $5$, $10$, and $45$ years from the fall of the last part of the clump. Time points are indicated in the upper right corner of each graph. }
\end{figure}

Let us consider in detail the model with perturbation parameters $r_0=3$~au and $dr=2$~au, $m_C=M_J$. The particles of the clump and the involved matter of the disc move towards the star along an elongated spiral (Fig.~\ref{fig:disk}). Because in the case $L<1$ the initial velocity of the clump particles less then Keplerian velocity. Thus, the particles start from the apocenter ($A_a=R$). It is easy to show that in the ballistic approximation the semi-major axis of the particle's orbit is $A=\frac{R}{2-L^2}$, and the eccentricity of the orbit is $e=1-L^2$, then the pericentric distance is $A_p=\frac{RL^2}{2-L^2}$. If $L=0.8$ the orbits of particles launched from a distance $R=3$~au have the semi-major axis $A\approx 2.2$~au, pericentric distance $A_p\approx 1.4$~au and period $P \approx 3.28$~yrs ($P\approx 7$~yrs for $R=5$~au). Thus, by the time $T=10$, when the spiral is twisted into a ring, the particles located near the inner boundary of the clump passes through the pericentre three times, which is consistent with the calculations.  

 In the gas-dynamic case, the particles can approach the star closer due to the viscosity of the disc. That is, the pericentric distance of the particles in the calculations is $a_p<A_p$. Therefore, for small $A_p$, the clump matter can approach the star several times and cause a series of bursts in the accretion rate.

\begin{figure}
\centering
\includegraphics[width=0.7\textwidth]{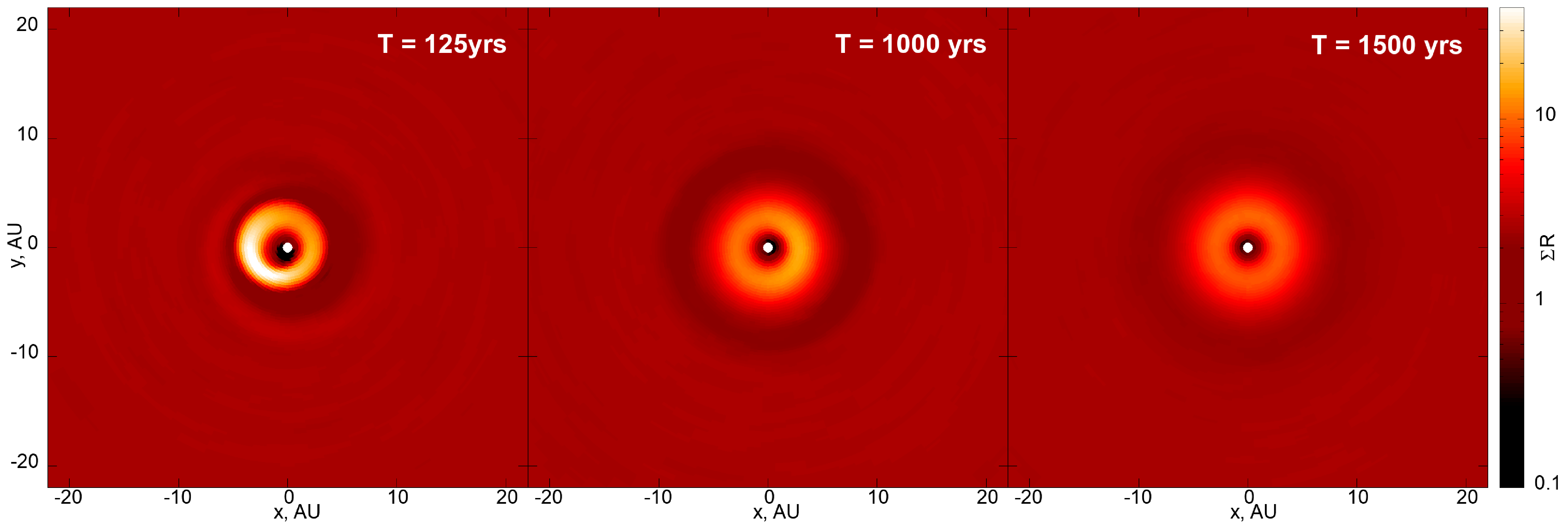}
\caption{\label{fig:sig20} The color scale shows of the surface density multiplied by the distance from the star after $125$, $1000$ and $1500$ years (indicated in the upper right corner) from the moment the clump appeared in calculated units. }
\end{figure}

Approximately $45$ years after the fall of the clump, the elongated spiral transforms into an asymmetric ring (Fig.~\ref{fig:disk}). The density of the ring decreases with time due to expansion. The particles of the ring move in quasi-elliptical orbits. A region of increased density is formed near the apocenter, since the particle velocities in this region are minimal. Under the action of viscous forces and self-gravity of the disk, the orbits of particles tend to decrease the eccentricity, and the position of the apocenter of the orbits changes.
Therefore the high-density region slowly move, completing in $1000$ years about half a revolution around the disk (Fig.~\ref{fig:sig20}). Azimuthal asymmetry in the ring at a distance of $R\approx2$~au  is present until the end of the calculations. In addition, a spiral density wave propagates across the disk (Fig.~\ref{fig:sig20}).
  

\begin{figure}
\centering
\includegraphics[width=0.7\textwidth]{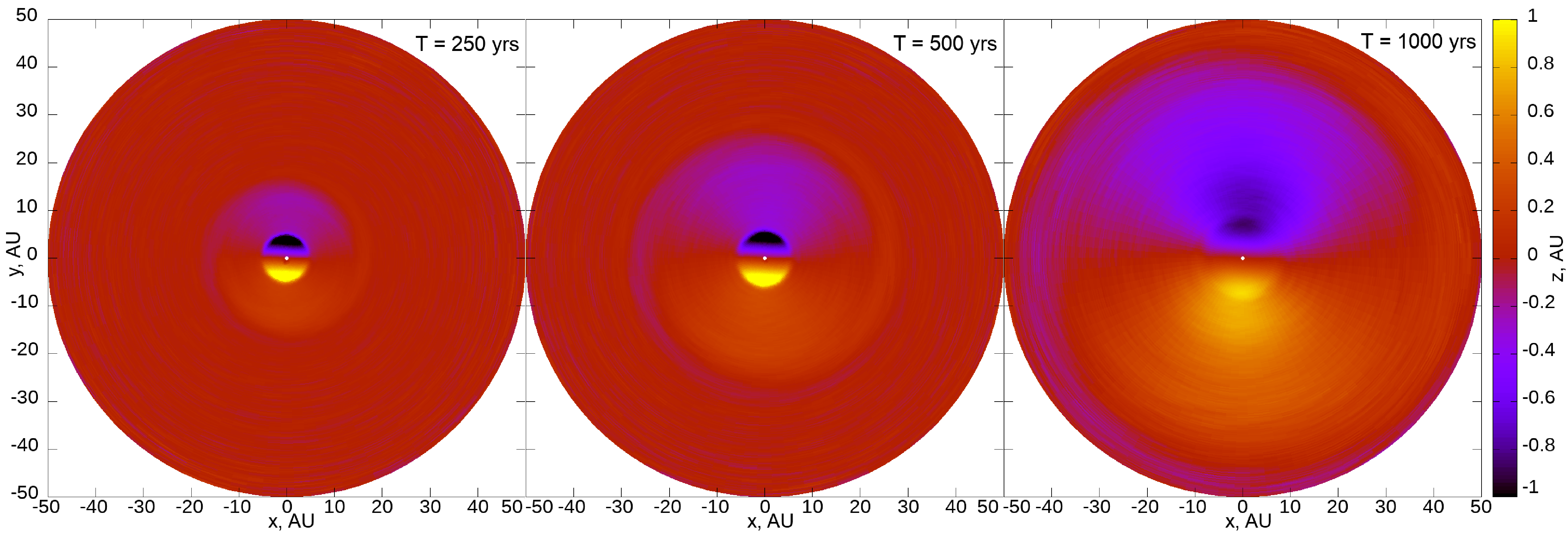}
\caption{\label{fig:inclIm} The color scale shows the average value of the $z$ coordinate in astronomical units in the cells described in the section~\ref{sec:method} for the time points of $250$, $500$ and $1000$ years (indicated in the upper right corner).}
\end{figure}

Since at the initial time the clump has a vertical velocity component, its particles drag the particles of the disk vertically. Therefore, the plane of the disk is noticeably distorted. At the initial stages of system relaxation, the inner part of the disk inclines, and over time, the distortion also affects the periphery of the disk. In this case, the maximum inclination of the disk plane relative to the initial position occurs near the $y$ axis (Fig.~\ref{fig:inclIm}). Along the $x$ axis, the disk retains its original position, since in the problem under consideration the orbital planes of the clump particles and the disk plane intersect near the $x$ axis.

\begin{figure}
\centering
\includegraphics[width=0.7\textwidth]{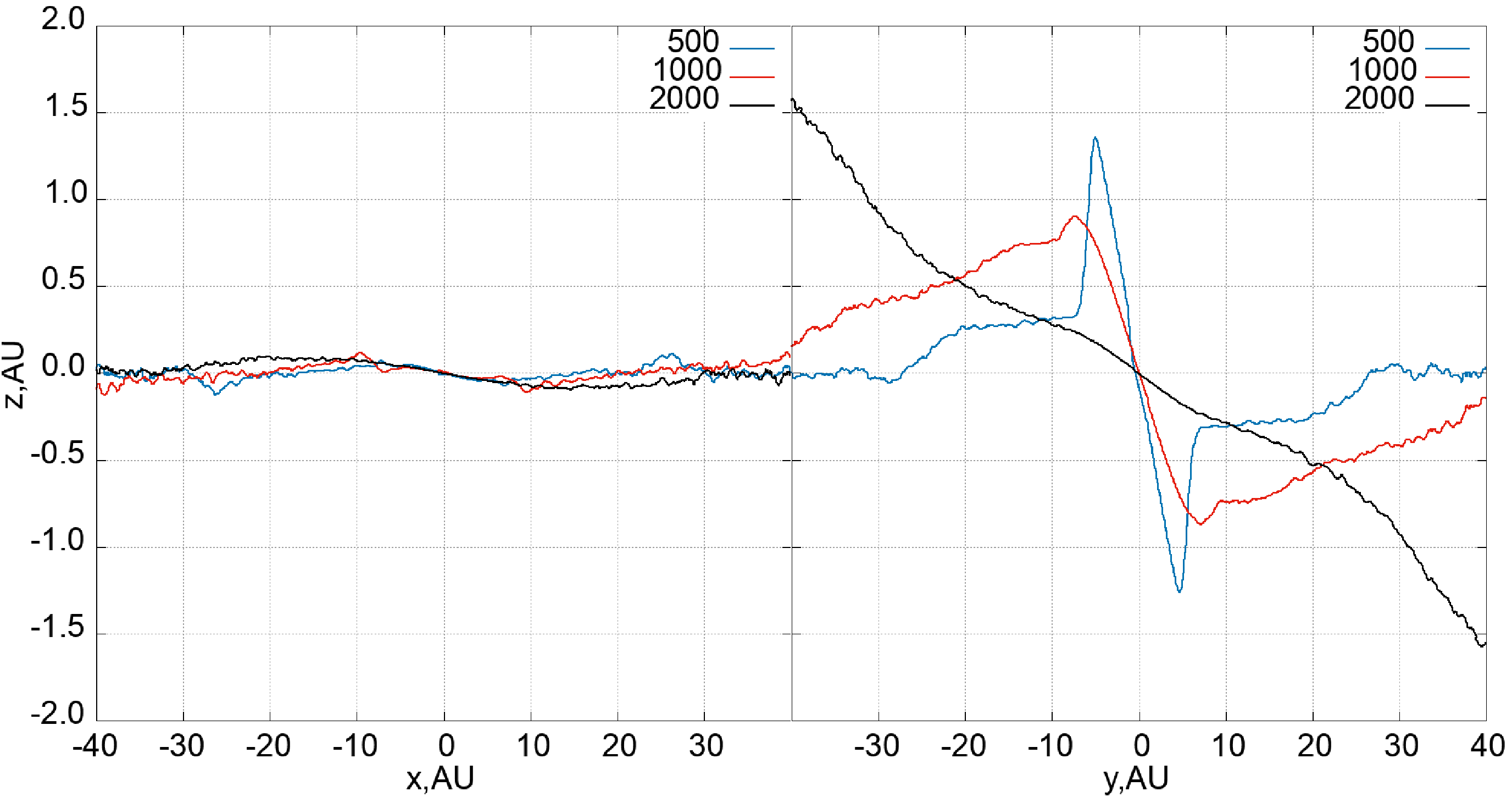}
\caption{\label{fig:inclXy} The average value of the $z$ coordinate along the $x$ (left) and $y$ (right) axes. The blue line corresponds to $500$ years, the red line corresponds to $1000$ years, and the black line corresponds to $2000$ years.}
\end{figure}

Fig.~\ref{fig:inclXy} shows the average value of the $z$ coordinate for sections along the $x$ (left) and $y$ (right) axes. It can be seen that the maximum inclination of the disk decreases with time, at the time of $500$ years it is $\sim 15^\circ$, by $1000$ years it is $\sim 7^\circ$ and by the end of the calculations in $2000$ years comes to $2^\circ$.

\subsection{Accretion rate burst}

The asymmetric dense ring is formed from the matter of the clump and the disk. It approaches to the accretion region in the process of relaxation. Approximately several tens years after the fall of the clump, the accretion rate begins to grow relative to its unperturbed value.

The unperturbed accretion rate inside a radius of $0.5$~au was $\sim 7\cdot 10^{-8}$ $M_\odot$ yr$^{-1}$. Calculations showed that the fall of a clump with a mass equal to $0.1~M_J$ leads to an increase in the accretion rate by about $1.3$ times, $0.3~M_J$ by $\sim 2.5$ times, $0.5~M_J$ by $\sim 3.7$ times, and $1~M_J$ by $\sim 7.7$ times. 

\begin{figure}
\centering
\includegraphics[width=0.7\textwidth]{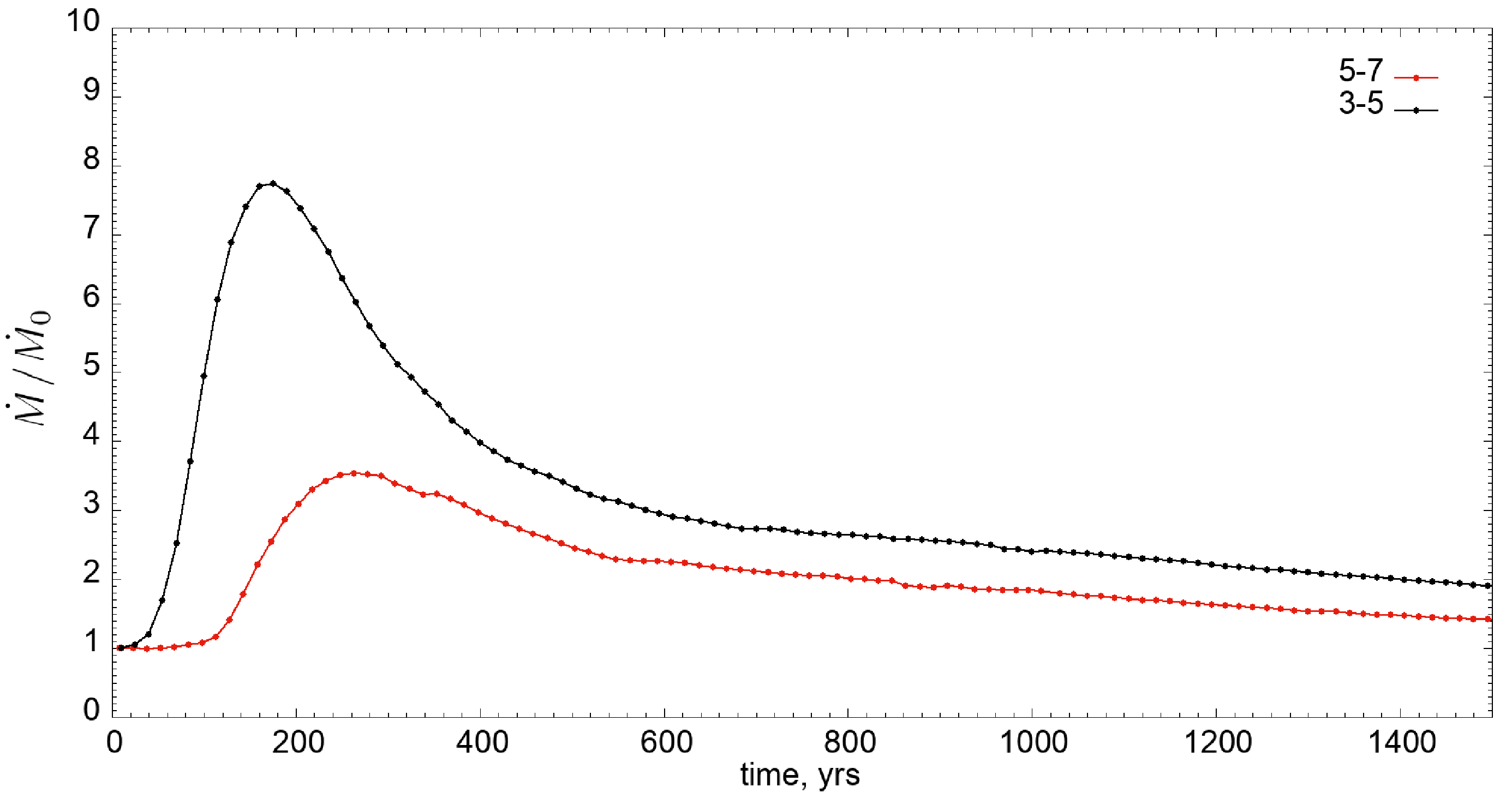}
\caption{\label{fig:acc} The ratio of the accretion rate onto the star (within a radius of $0.5$~au) to the undisturbed accretion rate as a function of time in years for models with $dr=2$~au, $m_C=M_J$, $L=0.8$, $I=30^\circ$. The black line shows the model with $r_0=3$~au, and the red one with $r_0=5$~au. Dependences are averaged over three points, and then smoothed over three points using the moving average method. }
\end{figure}

Fig.~\ref{fig:acc} shows the relative changes in the accretion rate from the undisturbed state for the case when the clump mass is $1 M_J$. Two models are presented, which differ in the position of the inner radius of the impact area of the clump: $R_0=3$~au and $R_0=5$~au. An analysis of the accretion rate showed that in the first case, the flare begins $25$~yrs after the fall of the clump, the accretion rate reaches a maximum in $135$~yrs, and the flare duration is $140$~yrs. In the second case, the flare begins in $105$~yrs after the fall of the clump, reaches a maximum after $150$~years and lasts $250$~yrs. 

By the time of $1500$ years, the accretion rate exceeds the initial value by about $2$ times. Displacement of the clump fall point by $R_0=5$~au increases the burst time to $100$ years, the maximum phase lasts $\sim 60$ years, while the accretion rate increases $\sim3.5$ times.

\begin{figure}
\centering
\includegraphics[width=0.7\textwidth]{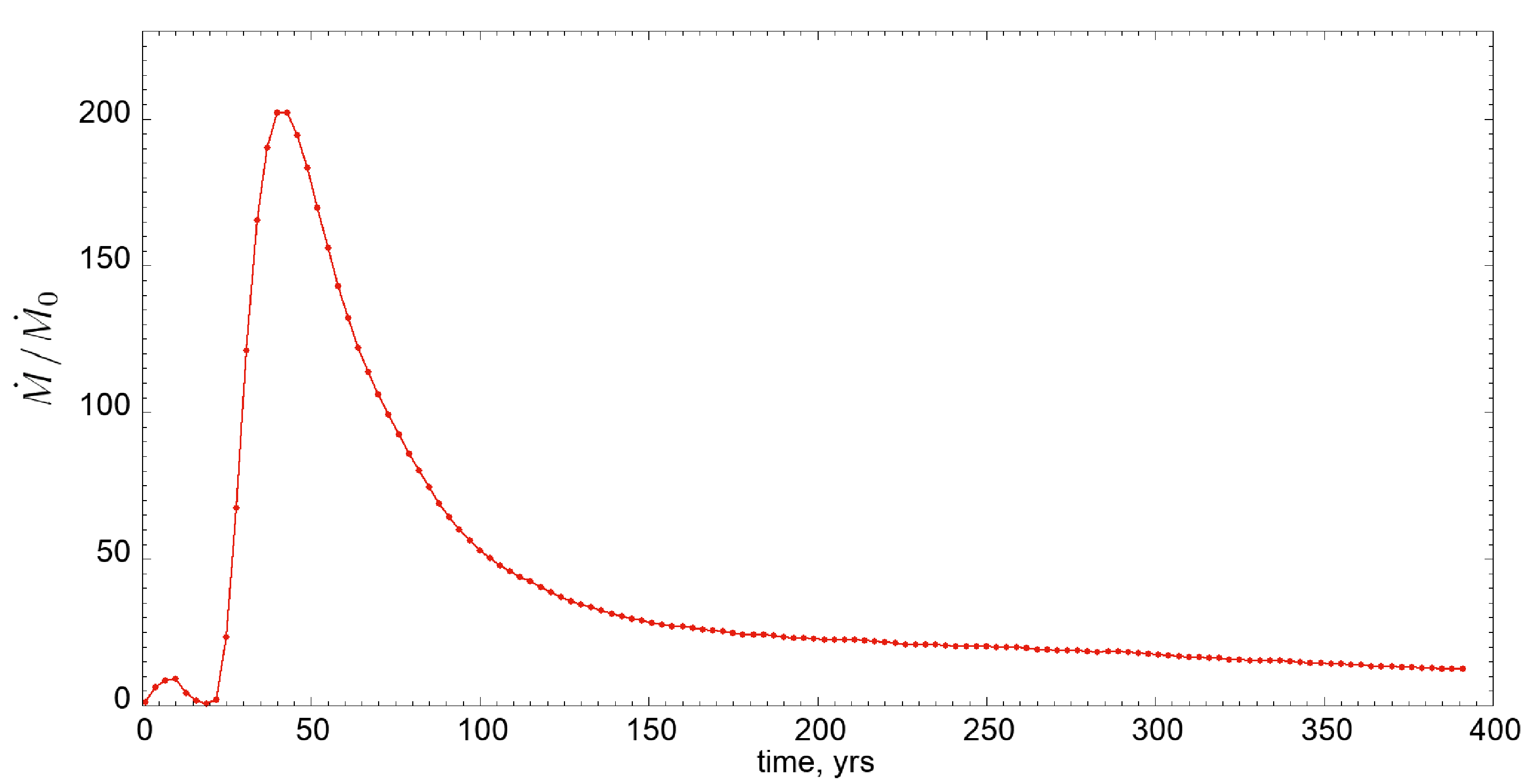}
\caption{\label{fig:fuor} Same as Fig.~\ref{fig:acc} for a model with $r_0=3$~au, $dr=1$~au, $m_C=3M_J$, $L=0.66$, $I=45^\circ$. }
\end{figure}

Fig.~\ref{fig:fuor} shows the most extreme model from our calculations with parameters: $r_0=3$~au, $dr=1$~au, $m_C=3M_J$, $L=0.66$, $I=45^\circ$. It can be seen that in this case the accretion rate increases $\sim 200$ times and reaches $1.4\cdot 10^{-5} M_\odot yr^{-1}$. The outburst reaches a maximum in 15 years, which is near the upper limit of the rise time of the accretion rate for known FUORs~\citep{2021A&A...647A..44V}. Flare duration is $\sim 30$ years and over a time of 50 years, the accretion rate decreases by $\sim 3$ times.

The calculations showed that with an increase in the perturbation mass, the inclination angle of the velocity vector $I$, and with a decrease in the parameter $L$, the accretion rate rise time decreases, and it increases to the value at the flare maximum. The displacement of the place where the clump fell to the star and the decrease in its radial dimensions affect in the same way due to the increase in the perturbation density. In addition, for smaller $r_0$ the burst starts earlier, as expected.

\subsection{Radiation of the disk}
\begin{figure}
\centering
\includegraphics[width=0.9\textwidth]{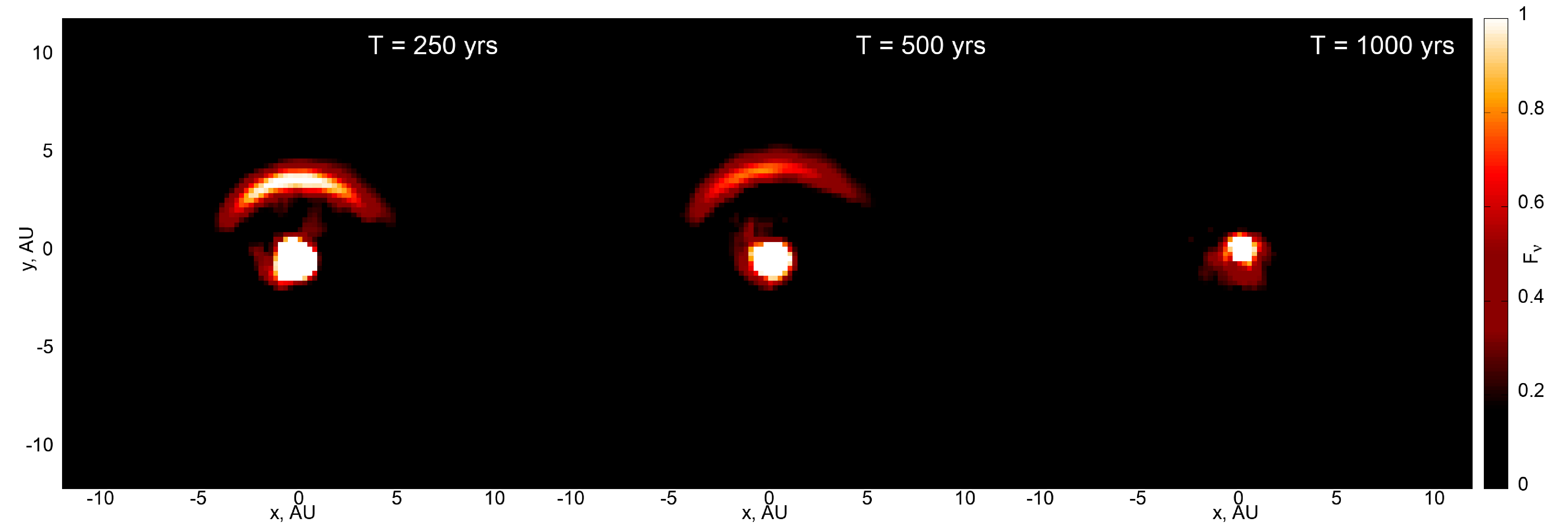}
\caption{\label{fig:inclIm3} Images of the disk at a wavelength of $3$~$\mu$m. The color scale shows the flux ($F_\nu$) in arbitrary units for the time points of $250$, $500$ and $1000$ years (indicated in the upper right corner).}
\end{figure}

Disk distortions that occur during the decay of the gas clump show themselves in the form of asymmetries on the image of the disk. In the near infrared (IR) area of the spectrum, the inclination of the internal part of the disk relative to the periphery do strong influence on the disk image. A matter towering above the original plane of the disk (near the negative part of the $y$ axis) casts a shadow to the outer layers of the disk. At the same time, from the opposite side (the positive part of the $ y $ axis), a dense ring-shaped structure is illuminated by the direct radiation of the star. Therefore, on the image of a disk at a wavelength of $3$~$\mu$m, a horseshoe shaped bright structure is visible in the semi-plate of positive values of the coordinates of $y$. In addition, near the star a distorted internal border of the disk radiates (Fig.~\ref{fig:inclIm3}).

\begin{figure}
\centering
\includegraphics[width=0.9\textwidth]{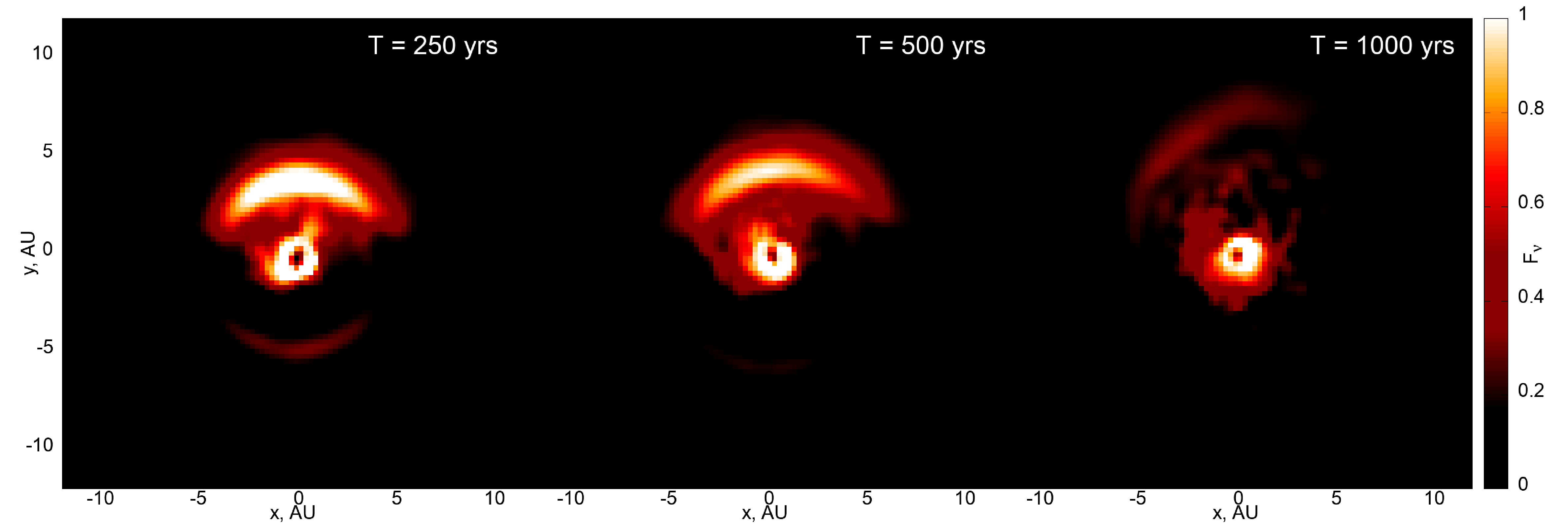}
\caption{\label{fig:inclIm100} Same as Fig.~\ref{fig:inclIm3} at $100$~$\mu$m.}
\end{figure}

At $100$~$\mu$m (far IR), the disk image also shows an asymmetry similar to that seen at $3$~$\mu$m. The illuminated area below the original plane of the disk also begins to radiate through the disk. Thus, the disk image is one bright horseshoe asymmetry and a weaker symmetrical inhomogeneity ~(Fig.~\ref{fig:inclIm100}).

\begin{figure}
\centering
\includegraphics[width=0.9\textwidth]{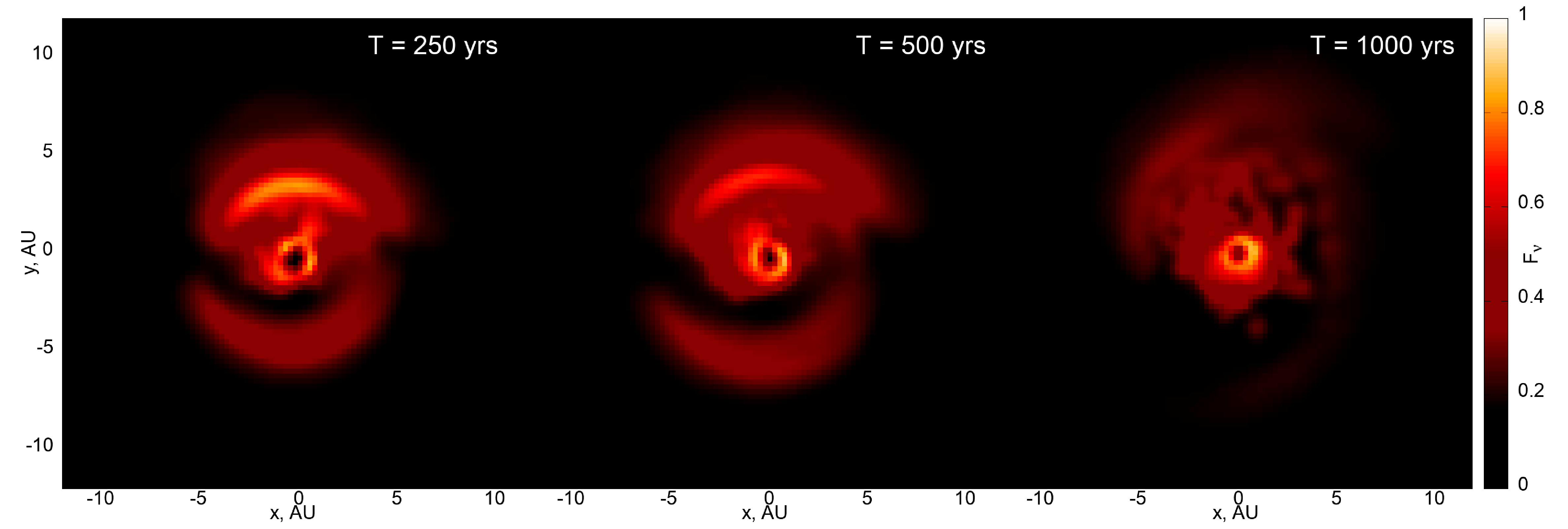}
\caption{\label{fig:inclIm1} Same as Fig.~\ref{fig:inclIm3} at $1$~mm.}
\end{figure}

In the millimeter region of the spectrum, the protoplanetary disk is transparent to radiation, so the image shows the hottest and densest regions, regardless of their position relative to the original disk plane. The image at a wavelength of $1$~mm for $\sim500$~years shows a bright structure with two symmetrical shadows along the $x$ axis. A line of contact between the tilted inner disk and the original plane of the disk runs along this axis. Therefore, these parts of the disk are less heated in comparison with the areas of the inner inclined disk that rise above the disk plane. At later stages, ragged asymmetry is visible in the central part of the disk, as well as areas of shadow (Fig.~\ref{fig:inclIm1}).

\section{Discussion}
The model calculations presented above show that the dynamic response of the circumstellar disk to the fall of a massive clump can lead to a powerful burst of accretion activity, which is characteristic of FUORs. The outburst of the accretion rate shown in Fig.~\ref{fig:fuor} resembles in its form the flare of the known FUORs FU Ori and V1057 Cyg with their fast rise and slow fall. Calculations have also shown that when the clump falls on the disk at a smaller distance from the star, the dynamic relaxation of the disk occurs faster (Fig.~\ref{fig:acc}).

The question arises: where do such massive clumps of gas come from in the immediate environment of a young star. The answer to this question is given by the results of modeling the development of gravitational instability in protoplanetary disks by~\citet{2005ApJ...633L.137V}. They showed that on the periphery of young disks, gravitational instability works very efficiently and leads to the formation of massive gas clumps, some of which migrate inside the disk, losing matter and angular momentum along the way. Some of the clumps are lost by the discs~\citep{2012ApJ...750...30B,2017A&A...608A.107V}. This process resembles a phase of the violent relaxation of young clusters~\citep[see e.g.][]{1998MNRAS.294...47G}. During this phase they lose some part of members and become more stable. A similar process is also realized in the early stages of the protoplanetary disks evolution. As a result, the star formation regions have to contain a lot of the clumps of gas and dust.

Indeed, such clumps are often observed as compact dark spots against the background of bright reﬂection nebulae and HII regions, and their number increases with decreasing size~\citep{2007AJ....133.1795G,2018Ap&SS.363...28G}. Their fall onto the disks, as shown by our calculations, can stimulate outbursts of accretion activity resembling FUOR ﬂares. This conclusion is in line with the hypothesis of~\citet{2015ApJ...805...15B} on the important role of clumpy accretion in maintaining intense accretion onto young stars.

Recently, \citet{2019A&A...628A..20D} and~\citet{2021A&A...656A.161K} computed 3D models of the gas-dynamic response of a circumstellar disk to the fall of a massive cloud. Similar calculations, taking into account the magnetic field of the disk and cloud, were carried out in~\citet{2022ApJ...941..154U}. In contrast to our paper, these authors considered the case when the fall of the cloudlet occurs at the periphery of the disk. Calculations have shown that, depending on the angle of incidence and the mass of the cloudlet, the outer (perturbed) region of the disk can get an inclination relative to the inner one. As in our case, in the process of relaxation, the inclinations of the inner and outer parts of the disk gradually level out. However, this is where the similarities between the two models end. In our case, the clump falls on the inner area of the disc, and it is this area that gets an inclination relative to the outer (main) part of the disc. In the models of~\citet{2021A&A...656A.161K} everything happens exactly the opposite: the inner disk is the primary disk formed at the birth of a star, the outer one is perturbed when a massive clump falls and get an inclination relative to the inner, primary disk. It is obvious that the appearance of a burst of accretion activity of a star, similar to that shown in Fig.~\ref{fig:fuor} is impossible in such models within a short time after the fall of the cloudlet. However, the late infall of a cloudlet can provoke the development of gravitational instability in disk. These processes leading to rapid transport of angular momentum of matter to the star and, accordingly, to accretion bursts.  

Observations of protoplanetary disks using the Atacama Large Millimeter/submillimeter Array
(ALMA) interferometer showed the presence of various types of structures on protoplanetary disks~\citep{2018A&A...619A.161C,2018ApJ...869L..43H, 2018ApJ...869L..42H,2018ApJ...869L..50P}. In particular, bright ring-shaped structures, spirals, vortex-like structures, as well as shadow areas are observed. In a recent paper~\citep{2021AJ....161...33V}, a detailed study of 16 disks with asymmetries of gaseous and dusty media was carried out. The authors note that for some objects there is evidence of the presence of a companion responsible for the formation of asymmetries in the disk. For some objects, signs of a companion of the stars are absent. The clumpy accretion model can explain the appearance of bright structures in the protoplanetary image (as was shown in Paper I), as well as the presence of quasi-symmetric shadows without assumptions about the presence of a companion. 
\newpage
\section{Conclusion}
Thus, the consequences of falling massive clumps on the disk that we have considered allow us to explain the following observed phenomena:
\begin{itemize}
\item Formation of structures (spirals, rings, etc.) observed in images of protoplanetary disks obtained with the ALMA (Paper I) interferometer. The lifetime of such structures depends on the place where the clump fell and its mass. In the models considered in Paper I, the lifetime of the rings was several thousand years.
\item Tilt of the inner area of the disk relative to the periphery. In our models, the inner disk with a radius on the order of several AU inclines after the fall of the clump relative to the outer (main) part of the disc. The characteristic relaxation time of the disk inclination angle is on the order of a thousand years. In the models~\citet{2019A&A...628A..20D, 2021A&A...656A.161K}, on the contrary, the periphery of the disk acquires an inclination. The alignment of the inclination angles of the outer and inner regions of the disk in this case is slower and may take several tens of thousands of years. Note that the problem of the origin of the inclination of the inner disk relative to the outer one first arose in the study of the circumstellar disk $\beta$ Pic and has since been actively discussed in the literature (see~\citet{2022A&A...658A.183B} and references there).

\item FOURs Flares. In order to create accretion bursts with characteristic development times similar to the observed FOUR flares, it is necessary that the  site of the massive clump impact on the disk be in the vicinity of a star with a radius of the order of several astronomical units. This requires that some of the clumps have orbits with a small specific angular momentum and large residual velocity vector inclination to the disk plane.
\end{itemize}

In the paper, we did not consider options for the fall of the clump at a large angle to the disk plane or with the opposite direction of the orbital spin, which are theoretically possible. In the latter case, the specific angular momentum of the perturbed region on the disk may turn out to be much less than the orbital one, as a result of which the perturbation may again fall on the disk, but much closer to the center. Such two-cascade processes can also lead to accretion bursts and deserve separate consideration.  

The process of clumpy accretion onto a protoplanetary disk, as well as the process of a close flyby of a star and the late infall of cloudlets, significantly distorts the structure of the disk. Under certain circumstances, all these processes can lead to the development of various instabilities. Including thermal, gravitational, magneto-rotational instabilities or a disk fragmentation. The consequence of such events may be create an outburst of the accretion onto the star with short rise time. In addition, these processes can contribute to the formation of exoplanets in exotic orbits: inclined to the disk plane, highly eccentric, and retrograde ones. 

{\bf Acknowledgments.} Simulations were partially performed using the resources of the Joint SuperComputer Center of the Russian Academy of Sciences — Branch of Federal State Institution ``Scientific Research Institute for System Analysis of the Russian Academy of Sciences''\footnote{\href{https://www.jscc.ru/}{https://www.jscc.ru/}}~\cite{2019LG..40..1835}.

\software{Gadget-2~\citep{2001NewA....6...79S,
2005MNRAS.364.1105S}, RADMC-3D~\cite{2012ascl.soft02015D}}

\bibliographystyle{aasjournal}
\bibliography{DG21}{}

\begin{thebibliography}{}
\expandafter\ifx\csname natexlab\endcsname\relax\def\natexlab#1{#1}\fi
\providecommand{\url}[1]{\href{#1}{#1}}
\providecommand{\dodoi}[1]{doi:~\href{http://doi.org/#1}{\nolinkurl{#1}}}
\providecommand{\doeprint}[1]{\href{http://ascl.net/#1}{\nolinkurl{http://ascl.net/#1}}}
\providecommand{\doarXiv}[1]{\href{https://arxiv.org/abs/#1}{\nolinkurl{https://arxiv.org/abs/#1}}}

\bibitem[{{Audard} {et~al.}(2014){Audard}, {{\'A}brah{\'a}m}, {Dunham},
  {Green}, {Grosso}, {Hamaguchi}, {Kastner}, {K{\'o}sp{\'a}l}, {Lodato},
  {Romanova}, {Skinner}, {Vorobyov}, \& {Zhu}}]{2014prpl.conf..387A}
{Audard}, M., {{\'A}brah{\'a}m}, P., {Dunham}, M.~M., {et~al.} 2014, in
  Protostars and Planets VI, ed. H.~{Beuther}, R.~S. {Klessen}, C.~P.
  {Dullemond}, \& T.~{Henning}, 387--410,
  \dodoi{10.2458/azu_uapress_9780816531240-ch017}

\bibitem[{{Bae} {et~al.}(2015){Bae}, {Hartmann}, \&
  {Zhu}}]{2015ApJ...805...15B}
{Bae}, J., {Hartmann}, L., \& {Zhu}, Z. 2015, \apj, 805, 15,
  \dodoi{10.1088/0004-637X/805/1/15}

\bibitem[{{Basu} \& {Vorobyov}(2012)}]{2012ApJ...750...30B}
{Basu}, S., \& {Vorobyov}, E.~I. 2012, \apj, 750, 30,
  \dodoi{10.1088/0004-637X/750/1/30}

\bibitem[{{Bohn} {et~al.}(2022){Bohn}, {Benisty}, {Perraut}, {van der Marel},
  {W{\"o}lfer}, {van Dishoeck}, {Facchini}, {Manara}, {Teague}, {Francis},
  {Berger}, {Garcia-Lopez}, {Ginski}, {Henning}, {Kenworthy}, {Kraus},
  {M{\'e}nard}, {M{\'e}rand}, \& {P{\'e}rez}}]{2022A&A...658A.183B}
{Bohn}, A.~J., {Benisty}, M., {Perraut}, K., {et~al.} 2022, \aap, 658, A183,
  \dodoi{10.1051/0004-6361/202142070}

\bibitem[{{Bohren} \& {Huffman}(1998)}]{1998asls.book.....B}
{Bohren}, C.~F., \& {Huffman}, D.~R. 1998, {Absorption and Scattering of Light
  by Small Particles}

\bibitem[{{Borchert} {et~al.}(2022{\natexlab{a}}){Borchert}, {Price}, {Pinte},
  \& {Cuello}}]{2022MNRAS.510L..37B}
{Borchert}, E. M.~A., {Price}, D.~J., {Pinte}, C., \& {Cuello}, N.
  2022{\natexlab{a}}, \mnras, 510, L37, \dodoi{10.1093/mnrasl/slab123}

\bibitem[{{Borchert} {et~al.}(2022{\natexlab{b}}){Borchert}, {Price}, {Pinte},
  \& {Cuello}}]{2022MNRAS.517.4436B}
---. 2022{\natexlab{b}}, \mnras, 517, 4436, \dodoi{10.1093/mnras/stac2872}

\bibitem[{{Cazzoletti} {et~al.}(2018){Cazzoletti}, {van Dishoeck}, {Pinilla},
  {Tazzari}, {Facchini}, {van der Marel}, {Benisty}, {Garufi}, \&
  {P{\'e}rez}}]{2018A&A...619A.161C}
{Cazzoletti}, P., {van Dishoeck}, E.~F., {Pinilla}, P., {et~al.} 2018, Astron.
  Astrophys., 619, A161, \dodoi{10.1051/0004-6361/201834006}

\bibitem[{{Chiang} \& {Goldreich}(1997)}]{1997ApJ...490..368C}
{Chiang}, E.~I., \& {Goldreich}, P. 1997, Astrophys. J., 490, 368,
  \dodoi{10.1086/304869}

\bibitem[{{Cuello} {et~al.}(2019){Cuello}, {Dipierro}, {Mentiplay}, {Price},
  {Pinte}, {Cuadra}, {Laibe}, {M{\'e}nard}, {Poblete}, \&
  {Montesinos}}]{2019MNRAS.483.4114C}
{Cuello}, N., {Dipierro}, G., {Mentiplay}, D., {et~al.} 2019, \mnras, 483,
  4114, \dodoi{10.1093/mnras/sty3325}

\bibitem[{{Cuello} {et~al.}(2020){Cuello}, {Louvet}, {Mentiplay}, {Pinte},
  {Price}, {Winter}, {Nealon}, {M{\'e}nard}, {Lodato}, {Dipierro},
  {Christiaens}, {Montesinos}, {Cuadra}, {Laibe}, {Cieza}, {Dong}, \&
  {Alexander}}]{2020MNRAS.491..504C}
{Cuello}, N., {Louvet}, F., {Mentiplay}, D., {et~al.} 2020, \mnras, 491, 504,
  \dodoi{10.1093/mnras/stz2938}

\bibitem[{{Demidova}(2016)}]{2016Ap.....59..449D}
{Demidova}, T.~V. 2016, Astrophysics, 59, 449,
  \dodoi{10.1007/s10511-016-9448-3}

\bibitem[{{Demidova} \& {Grinin}(2022)}]{2022ApJ...930..111D}
{Demidova}, T.~V., \& {Grinin}, V.~P. 2022, \apj, 930, 111,
  \dodoi{10.3847/1538-4357/ac53a6}

\bibitem[{{Dohnanyi}(1969)}]{1969JGR....74.2531D}
{Dohnanyi}, J.~S. 1969, \jgr, 74, 2531, \dodoi{10.1029/JB074i010p02531}

\bibitem[{{Dorschner} {et~al.}(1995){Dorschner}, {Begemann}, {Henning},
  {Jaeger}, \& {Mutschke}}]{1995A&A...300..503D}
{Dorschner}, J., {Begemann}, B., {Henning}, T., {Jaeger}, C., \& {Mutschke}, H.
  1995, Astron. Astrophys., 300, 503

\bibitem[{{Dullemond} \& {Dominik}(2004)}]{2004A&A...421.1075D}
{Dullemond}, C.~P., \& {Dominik}, C. 2004, Astron. Astrophys., 421, 1075,
  \dodoi{10.1051/0004-6361:20040284}

\bibitem[{{Dullemond} {et~al.}(2012){Dullemond}, {Juhasz}, {Pohl}, {Sereshti},
  {Shetty}, {Peters}, {Commercon}, \& {Flock}}]{2012ascl.soft02015D}
{Dullemond}, C.~P., {Juhasz}, A., {Pohl}, A., {et~al.} 2012, {RADMC-3D: A
  multi-purpose radiative transfer tool}.
\newblock \doeprint{1202.015}

\bibitem[{{Dullemond} {et~al.}(2019){Dullemond}, {K{\"u}ffmeier}, {Goicovic},
  {Fukagawa}, {Oehl}, \& {Kramer}}]{2019A&A...628A..20D}
{Dullemond}, C.~P., {K{\"u}ffmeier}, M., {Goicovic}, F., {et~al.} 2019, \aap,
  628, A20, \dodoi{10.1051/0004-6361/201832632}

\bibitem[{{Dutrey} {et~al.}(1994){Dutrey}, {Guilloteau}, \&
  {Simon}}]{1994A&A...286..149D}
{Dutrey}, A., {Guilloteau}, S., \& {Simon}, M. 1994, Astron. Astrophys., 286,
  149

\bibitem[{{Fischer} {et~al.}(2022){Fischer}, {Hillenbrand}, {Herczeg},
  {Johnstone}, {K{\'o}sp{\'a}l}, \& {Dunham}}]{2022arXiv220311257F}
{Fischer}, W.~J., {Hillenbrand}, L.~A., {Herczeg}, G.~J., {et~al.} 2022, arXiv
  e-prints, arXiv:2203.11257, \dodoi{10.48550/arXiv.2203.11257}

\bibitem[{{Forgan} \& {Rice}(2010)}]{2010MNRAS.402.1349F}
{Forgan}, D., \& {Rice}, K. 2010, \mnras, 402, 1349,
  \dodoi{10.1111/j.1365-2966.2009.15974.x}

\bibitem[{{Gahm} {et~al.}(2007){Gahm}, {Grenman}, {Fredriksson}, \&
  {Kristen}}]{2007AJ....133.1795G}
{Gahm}, G.~F., {Grenman}, T., {Fredriksson}, S., \& {Kristen}, H. 2007, \aj,
  133, 1795, \dodoi{10.1086/512036}

\bibitem[{{Goodwin}(1998)}]{1998MNRAS.294...47G}
{Goodwin}, S.~P. 1998, \mnras, 294, 47,
  \dodoi{10.1046/j.1365-8711.1998.01192.x}

\bibitem[{{Grenman} {et~al.}(2018){Grenman}, {Elfgren}, \&
  {Weber}}]{2018Ap&SS.363...28G}
{Grenman}, T., {Elfgren}, E., \& {Weber}, H. 2018, \apss, 363, 28,
  \dodoi{10.1007/s10509-017-3233-6}

\bibitem[{{Hartmann} \& {Kenyon}(1985)}]{1985ApJ...299..462H}
{Hartmann}, L., \& {Kenyon}, S.~J. 1985, \apj, 299, 462, \dodoi{10.1086/163713}

\bibitem[{{Hartmann} \& {Kenyon}(1996)}]{1996ARA&A..34..207H}
---. 1996, \araa, 34, 207, \dodoi{10.1146/annurev.astro.34.1.207}

\bibitem[{{Herbig}(1977)}]{1977ApJ...217..693H}
{Herbig}, G.~H. 1977, \apj, 217, 693, \dodoi{10.1086/155615}

\bibitem[{{Huang} {et~al.}(2018{\natexlab{a}}){Huang}, {Andrews}, {P{\'e}rez},
  {Zhu}, {Dullemond}, {Isella}, {Benisty}, {Bai}, {Birnstiel}, {Carpenter},
  {Guzm{\'a}n}, {Hughes}, {{\"O}berg}, {Ricci}, {Wilner}, \&
  {Zhang}}]{2018ApJ...869L..43H}
{Huang}, J., {Andrews}, S.~M., {P{\'e}rez}, L.~M., {et~al.} 2018{\natexlab{a}},
  Astrophys. J. Lett., 869, L43, \dodoi{10.3847/2041-8213/aaf7a0}

\bibitem[{{Huang} {et~al.}(2018{\natexlab{b}}){Huang}, {Andrews}, {Dullemond},
  {Isella}, {P{\'e}rez}, {Guzm{\'a}n}, {{\"O}berg}, {Zhu}, {Zhang}, {Bai},
  {Benisty}, {Birnstiel}, {Carpenter}, {Hughes}, {Ricci}, {Weaver}, \&
  {Wilner}}]{2018ApJ...869L..42H}
{Huang}, J., {Andrews}, S.~M., {Dullemond}, C.~P., {et~al.} 2018{\natexlab{b}},
  \apjl, 869, L42, \dodoi{10.3847/2041-8213/aaf740}

\bibitem[{{Jensen} \& {Haugb{\o}lle}(2018)}]{2018MNRAS.474.1176J}
{Jensen}, S.~S., \& {Haugb{\o}lle}, T. 2018, \mnras, 474, 1176,
  \dodoi{10.1093/mnras/stx2844}

\bibitem[{{Kopatskaya} {et~al.}(2013){Kopatskaya}, {Kolotilov}, \&
  {Arkharov}}]{2013MNRAS.434...38K}
{Kopatskaya}, E.~N., {Kolotilov}, E.~A., \& {Arkharov}, A.~A. 2013, \mnras,
  434, 38, \dodoi{10.1093/mnras/stt963}

\bibitem[{{Kuffmeier} {et~al.}(2021){Kuffmeier}, {Dullemond}, {Reissl}, \&
  {Goicovic}}]{2021A&A...656A.161K}
{Kuffmeier}, M., {Dullemond}, C.~P., {Reissl}, S., \& {Goicovic}, F.~G. 2021,
  \aap, 656, A161, \dodoi{10.1051/0004-6361/202039614}

\bibitem[{{Kuffmeier} {et~al.}(2018){Kuffmeier}, {Frimann}, {Jensen}, \&
  {Haugb{\o}lle}}]{2018MNRAS.475.2642K}
{Kuffmeier}, M., {Frimann}, S., {Jensen}, S.~S., \& {Haugb{\o}lle}, T. 2018,
  \mnras, 475, 2642, \dodoi{10.1093/mnras/sty024}

\bibitem[{{Labdon} {et~al.}(2021){Labdon}, {Kraus}, {Davies}, {Kreplin},
  {Monnier}, {Le Bouquin}, {Anugu}, {ten Brummelaar}, {Setterholm}, {Gardner},
  {Ennis}, {Lanthermann}, {Schaefer}, \& {Laws}}]{2021A&A...646A.102L}
{Labdon}, A., {Kraus}, S., {Davies}, C.~L., {et~al.} 2021, \aap, 646, A102,
  \dodoi{10.1051/0004-6361/202039370}

\bibitem[{{Malygin} {et~al.}(2017){Malygin}, {Klahr}, {Semenov}, {Henning}, \&
  {Dullemond}}]{2017A&A...605A..30M}
{Malygin}, M.~G., {Klahr}, H., {Semenov}, D., {Henning}, T., \& {Dullemond},
  C.~P. 2017, \aap, 605, A30, \dodoi{10.1051/0004-6361/201629933}

\bibitem[{{Offner} \& {McKee}(2011)}]{2011ApJ...736...53O}
{Offner}, S. S.~R., \& {McKee}, C.~F. 2011, \apj, 736, 53,
  \dodoi{10.1088/0004-637X/736/1/53}

\bibitem[{{Padoan} {et~al.}(2014){Padoan}, {Haugb{\o}lle}, \&
  {Nordlund}}]{2014ApJ...797...32P}
{Padoan}, P., {Haugb{\o}lle}, T., \& {Nordlund}, {\r{A}}. 2014, \apj, 797, 32,
  \dodoi{10.1088/0004-637X/797/1/32}

\bibitem[{{P{\'e}rez} {et~al.}(2018){P{\'e}rez}, {Benisty}, {Andrews},
  {Isella}, {Dullemond}, {Huang}, {Kurtovic}, {Guzm{\'a}n}, {Zhu}, {Birnstiel},
  {Zhang}, {Carpenter}, {Wilner}, {Ricci}, {Bai}, {Weaver}, \&
  {{\"O}berg}}]{2018ApJ...869L..50P}
{P{\'e}rez}, L.~M., {Benisty}, M., {Andrews}, S.~M., {et~al.} 2018, Astrophys.
  J. Lett., 869, L50, \dodoi{10.3847/2041-8213/aaf745}

\bibitem[{{Pfalzner}(2008)}]{2008A&A...492..735P}
{Pfalzner}, S. 2008, \aap, 492, 735, \dodoi{10.1051/0004-6361:200810879}

\bibitem[{{Savin} {et~al.}(2019){Savin}, {Shabanov}, {Telegin}, \&
  {Baranov}}]{2019LG..40..1835}
{Savin}, G., {Shabanov}, B., {Telegin}, P., \& {Baranov}, A. 2019, Lobachevskii
  Journal of Mathematics, 40, 1853, \dodoi{10.1134/S1995080219110271}

\bibitem[{{Springel}(2005)}]{2005MNRAS.364.1105S}
{Springel}, V. 2005, MNRAS, 364, 1105, \dodoi{10.1111/j.1365-2966.2005.09655.x}

\bibitem[{{Springel} {et~al.}(2001){Springel}, {Yoshida}, \&
  {White}}]{2001NewA....6...79S}
{Springel}, V., {Yoshida}, N., \& {White}, S. D.~M. 2001, New Astronomy, 6, 79,
  \dodoi{10.1016/S1384-1076(01)00042-2}

\bibitem[{{Szab{\'o}} {et~al.}(2021){Szab{\'o}}, {K{\'o}sp{\'a}l},
  {{\'A}brah{\'a}m}, {Park}, {Siwak}, {Green}, {Mo{\'o}r}, {P{\'a}l},
  {Acosta-Pulido}, {Lee}, {Cseh}, {Cs{\"o}rnyei}, {Hanyecz},
  {K{\"o}nyves-T{\'o}th}, {Krezinger}, {Kriskovics}, {Ordasi}, {S{\'a}rneczky},
  {Seli}, {Szak{\'a}ts}, {Szing}, \& {Vida}}]{2021ApJ...917...80S}
{Szab{\'o}}, Z.~M., {K{\'o}sp{\'a}l}, {\'A}., {{\'A}brah{\'a}m}, P., {et~al.}
  2021, \apj, 917, 80, \dodoi{10.3847/1538-4357/ac04b3}

\bibitem[{{Unno} {et~al.}(2022){Unno}, {Hanawa}, \&
  {Takasao}}]{2022ApJ...941..154U}
{Unno}, M., {Hanawa}, T., \& {Takasao}, S. 2022, \apj, 941, 154,
  \dodoi{10.3847/1538-4357/aca410}

\bibitem[{{van der Marel} {et~al.}(2021){van der Marel}, {Birnstiel}, {Garufi},
  {Ragusa}, {Christiaens}, {Price}, {Sallum}, {Muley}, {Francis}, \&
  {Dong}}]{2021AJ....161...33V}
{van der Marel}, N., {Birnstiel}, T., {Garufi}, A., {et~al.} 2021, \aj, 161,
  33, \dodoi{10.3847/1538-3881/abc3ba}

\bibitem[{{Vorobyov} \& {Basu}(2005)}]{2005ApJ...633L.137V}
{Vorobyov}, E.~I., \& {Basu}, S. 2005, \apjl, 633, L137, \dodoi{10.1086/498303}

\bibitem[{{Vorobyov} {et~al.}(2021){Vorobyov}, {Elbakyan}, {Liu}, \&
  {Takami}}]{2021A&A...647A..44V}
{Vorobyov}, E.~I., {Elbakyan}, V.~G., {Liu}, H.~B., \& {Takami}, M. 2021, \aap,
  647, A44, \dodoi{10.1051/0004-6361/202039391}

\bibitem[{{Vorobyov} {et~al.}(2017){Vorobyov}, {Steinrueck}, {Elbakyan}, \&
  {Guedel}}]{2017A&A...608A.107V}
{Vorobyov}, E.~I., {Steinrueck}, M.~E., {Elbakyan}, V., \& {Guedel}, M. 2017,
  \aap, 608, A107, \dodoi{10.1051/0004-6361/201731565}

\end{thebibliography}
\end{document}